\pdfoutput=1

\documentclass[12pt,a4paper]{article}

\usepackage{setspace}
\usepackage{slashbox}
\usepackage{ifthen} 
\newboolean{pdflatex}
\setboolean{pdflatex}{true} 

\newboolean{articletitles}
\setboolean{articletitles}{true} 

\newboolean{uprightparticles}
\setboolean{uprightparticles}{false} 

\newboolean{inbibliography}
\setboolean{inbibliography}{false} 

\usepackage[top=1in, bottom=1.25in, left=1in, right=1in]{geometry}

\usepackage{microtype}
\usepackage{lineno}  
\usepackage{xspace} 
\usepackage{caption} 

\usepackage{graphicx}  
\usepackage{color}
\usepackage{colortbl}
\graphicspath{{./figs/}} 

\usepackage{amsmath} 
\usepackage{amssymb}
\usepackage{amsfonts}
\usepackage{upgreek} 

\newcommand*\patchAmsMathEnvironmentForLineno[1]{%
\expandafter\let\csname old#1\expandafter\endcsname\csname #1\endcsname
\expandafter\let\csname oldend#1\expandafter\endcsname\csname
end#1\endcsname
 \renewenvironment{#1}%
   {\linenomath\csname old#1\endcsname}%
   {\csname oldend#1\endcsname\endlinenomath}%
}
\newcommand*\patchBothAmsMathEnvironmentsForLineno[1]{%
  \patchAmsMathEnvironmentForLineno{#1}%
  \patchAmsMathEnvironmentForLineno{#1*}%
}
\AtBeginDocument{%
\patchBothAmsMathEnvironmentsForLineno{equation}%
\patchBothAmsMathEnvironmentsForLineno{align}%
\patchBothAmsMathEnvironmentsForLineno{flalign}%
\patchBothAmsMathEnvironmentsForLineno{alignat}%
\patchBothAmsMathEnvironmentsForLineno{gather}%
\patchBothAmsMathEnvironmentsForLineno{multline}%
\patchBothAmsMathEnvironmentsForLineno{eqnarray}%
}

\usepackage{hyperref}    
\usepackage{cleveref}

\usepackage[all]{hypcap} 

\usepackage{xspace} 
\usepackage{upgreek}

\def\lhcb {\mbox{LHCb}\xspace}

\def\velo   {VELO\xspace}

\def\MagUp {\mbox{\em Mag\kern -0.05em Up}\xspace}

\ifthenelse{\boolean{uprightparticles}}%
{

 \def\Pmu         {\ensuremath{\upmu}\xspace}

 \def\Ppsi        {\ensuremath{\uppsi}\xspace}

 \def\PDelta      {\ensuremath{\Delta}\xspace}                 
 \def\PXi      {\ensuremath{\Xi}\xspace}                 
 \def\PLambda      {\ensuremath{\Lambda}\xspace}                 
 \def\PSigma      {\ensuremath{\Sigma}\xspace}                 
 \def\POmega      {\ensuremath{\Omega}\xspace}                 
 \def\PUpsilon      {\ensuremath{\Upsilon}\xspace}                 
 
 \def\PB      {\ensuremath{\mathrm{B}}\xspace}                 
                  
 \def\PD      {\ensuremath{\mathrm{D}}\xspace}

 \def\PH      {\ensuremath{\mathrm{H}}\xspace}                 
                  
 \def\PJ      {\ensuremath{\mathrm{J}}\xspace}                 
 \def\PK      {\ensuremath{\mathrm{K}}\xspace}

 \def\PW      {\ensuremath{\mathrm{W}}\xspace}

 \def\PZ      {\ensuremath{\mathrm{Z}}\xspace}                 
                  
 \def\Pb      {\ensuremath{\mathrm{b}}\xspace}                 
 \def\Pc      {\ensuremath{\mathrm{c}}\xspace}

 \def\Pi      {\ensuremath{\mathrm{i}}\xspace}

 \def\Pt      {\ensuremath{\mathrm{t}}\xspace}

}
{

 \def\Pmu         {\ensuremath{\mu}\xspace}

 \def\Ppsi        {\ensuremath{\psi}\xspace}                 
                  
 \mathchardef\PDelta="7101
 \mathchardef\PXi="7104
 \mathchardef\PLambda="7103
 \mathchardef\PSigma="7106
 \mathchardef\POmega="710A
 \mathchardef\PUpsilon="7107
                  
 \def\PB      {\ensuremath{B}\xspace}                 
                  
 \def\PD      {\ensuremath{D}\xspace}

 \def\PH      {\ensuremath{H}\xspace}                 
                  
 \def\PJ      {\ensuremath{J}\xspace}                 
 \def\PK      {\ensuremath{K}\xspace}

 \def\PW      {\ensuremath{W}\xspace}

 \def\PZ      {\ensuremath{Z}\xspace}                 
                  
 \def\Pb      {\ensuremath{b}\xspace}                 
 \def\Pc      {\ensuremath{c}\xspace}

 \def\Pi      {\ensuremath{i}\xspace}

 \def\Pt      {\ensuremath{t}\xspace}

}

\makeatletter
\ifcase \@ptsize \relax
  \newcommand{\miniscule}{\@setfontsize\miniscule{4}{5}}
\or
  \newcommand{\miniscule}{\@setfontsize\miniscule{5}{6}}
\or
  \newcommand{\miniscule}{\@setfontsize\miniscule{5}{6}}
\fi
\makeatother

\DeclareRobustCommand{\optbar}[1]{\shortstack{{\miniscule (\rule[.5ex]{1.25em}{.18mm})}
  \\ [-.7ex] $#1$}}

\def\mumu       {{\ensuremath{\Pmu^+\Pmu^-}}\xspace}

\def\H      {{\ensuremath{\PH^0}}\xspace}

\def\W      {{\ensuremath{\PW}}\xspace}

\def\Z      {{\ensuremath{\PZ}}\xspace}

\def\cquark    {{\ensuremath{\Pc}}\xspace}
\def\cquarkbar {{\ensuremath{\overline \cquark}}\xspace}
\def\ccbar     {{\ensuremath{\cquark\cquarkbar}}\xspace}
\def\bquark    {{\ensuremath{\Pb}}\xspace}
\def\bquarkbar {{\ensuremath{\overline \bquark}}\xspace}
\def\bbbar     {{\ensuremath{\bquark\bquarkbar}}\xspace}
\def\tquark    {{\ensuremath{\Pt}}\xspace}
\def\tquarkbar {{\ensuremath{\overline \tquark}}\xspace}
\def\ttbar     {{\ensuremath{\tquark\tquarkbar}}\xspace}

\def\kaon    {{\ensuremath{\PK}}\xspace}
  \def\Kbar    {{\kern 0.2em\overline{\kern -0.2em \PK}{}}\xspace}

\def\KorKbar    {\kern 0.18em\optbar{\kern -0.18em K}{}\xspace}

\def\Kstarz  {{\ensuremath{\kaon^{*0}}}\xspace}

  \def\Dbar    {{\kern 0.2em\overline{\kern -0.2em \PD}{}}\xspace}

\def\DorDbar    {\kern 0.18em\optbar{\kern -0.18em D}{}\xspace}

\def\Bbar    {{\ensuremath{\kern 0.18em\overline{\kern -0.18em \PB}{}}}\xspace}

\def\BorBbar    {\kern 0.18em\optbar{\kern -0.18em B}{}\xspace}

\def\jpsi     {{\ensuremath{{\PJ\mskip -3mu/\mskip -2mu\Ppsi\mskip 2mu}}}\xspace}

  \def\Y#1S{\ensuremath{\PUpsilon{(#1S)}}\xspace}

\def\Lbar        {{\ensuremath{\kern 0.1em\overline{\kern -0.1em\PLambda}}}\xspace}
\def\LorLbar    {\kern 0.18em\optbar{\kern -0.18em \PLambda}{}\xspace}

\newcommand{\decay}[2]{\ensuremath{#1\!\to #2}\xspace}         

\def\to                 {\ensuremath{\rightarrow}\xspace}

\def\AT#1     {\ensuremath{A_{\mathrm{T}}^{#1}}\xspace}           

\def\C#1      {\ensuremath{\mathcal{C}_{#1}}\xspace}                       
\def\Cp#1     {\ensuremath{\mathcal{C}_{#1}^{'}}\xspace}                    
\def\Ceff#1   {\ensuremath{\mathcal{C}_{#1}^{\mathrm{(eff)}}}\xspace}        
\def\Cpeff#1  {\ensuremath{\mathcal{C}_{#1}^{'\mathrm{(eff)}}}\xspace}       
\def\Ope#1    {\ensuremath{\mathcal{O}_{#1}}\xspace}                       
\def\Opep#1   {\ensuremath{\mathcal{O}_{#1}^{'}}\xspace}                    

\newcommand{\tev}{\ifthenelse{\boolean{inbibliography}}{\ensuremath{~T\kern -0.05em eV}\xspace}{\ensuremath{\mathrm{\,Te\kern -0.1em V}}}\xspace}
\newcommand{\gev}{\ensuremath{\mathrm{\,Ge\kern -0.1em V}}\xspace}
\newcommand{\mev}{\ensuremath{\mathrm{\,Me\kern -0.1em V}}\xspace}
\newcommand{\kev}{\ensuremath{\mathrm{\,ke\kern -0.1em V}}\xspace}
\newcommand{\ev}{\ensuremath{\mathrm{\,e\kern -0.1em V}}\xspace}
\newcommand{\gevc}{\ensuremath{{\mathrm{\,Ge\kern -0.1em V\!/}c}}\xspace}
\newcommand{\mevc}{\ensuremath{{\mathrm{\,Me\kern -0.1em V\!/}c}}\xspace}
\newcommand{\gevcc}{\ensuremath{{\mathrm{\,Ge\kern -0.1em V\!/}c^2}}\xspace}
\newcommand{\gevgevcccc}{\ensuremath{{\mathrm{\,Ge\kern -0.1em V^2\!/}c^4}}\xspace}
\newcommand{\mevcc}{\ensuremath{{\mathrm{\,Me\kern -0.1em V\!/}c^2}}\xspace}

\def\m    {\ensuremath{\mathrm{ \,m}}\xspace}

\def\cm   {\ensuremath{\mathrm{ \,cm}}\xspace}

\def\mm   {\ensuremath{\mathrm{ \,mm}}\xspace}

\def\mum  {\ensuremath{{\,\upmu\mathrm{m}}}\xspace}

\def\pb {\ensuremath{\mathrm{ \,pb}}\xspace}

\def\invfb   {\ensuremath{\mbox{\,fb}^{-1}}\xspace}

\def\ps   {\ensuremath{{\mathrm{ \,ps}}}\xspace}

\newcommand{\chisqndf}{\ensuremath{\chi^2/\mathrm{ndf}}\xspace}

\def\gsim{{~\raise.15em\hbox{$>$}\kern-.85em
          \lower.35em\hbox{$\sim$}~}\xspace}
\def\lsim{{~\raise.15em\hbox{$<$}\kern-.85em
          \lower.35em\hbox{$\sim$}~}\xspace}

\def\sqs   {\ensuremath{\protect\sqrt{s}}\xspace}

\def\ptot       {\mbox{$p$}\xspace}
\def\pt         {\mbox{$p_{\mathrm{ T}}$}\xspace}

\def\geant      {\mbox{\textsc{Geant4}}\xspace}

\def\pythia     {\mbox{\textsc{Pythia}}\xspace}

\def\tell1  {TELL1\xspace}
\def\ukl1   {UKL1\xspace}

\newcommand{\eg}{\mbox{\itshape e.g.}\xspace}
\newcommand{\ie}{\mbox{\itshape i.e.}\xspace}

\usepackage{cite} 
\usepackage{mciteplus}

\usepackage{longtable} 
\usepackage{subfigure}

\begin{document}

\renewcommand{\thefootnote}{\fnsymbol{footnote}}
\setcounter{footnote}{1}


\begin{titlepage}
\pagenumbering{roman}

\vspace*{-1.5cm}
\centerline{\large EUROPEAN ORGANIZATION FOR NUCLEAR RESEARCH (CERN)}
\vspace*{1.5cm}
\noindent
\begin{tabular*}{\linewidth}{lc@{\extracolsep{\fill}}r@{\extracolsep{0pt}}}
\ifthenelse{\boolean{pdflatex}}
{\vspace*{-2.7cm}\mbox{\!\!\!\includegraphics[width=.14\textwidth]{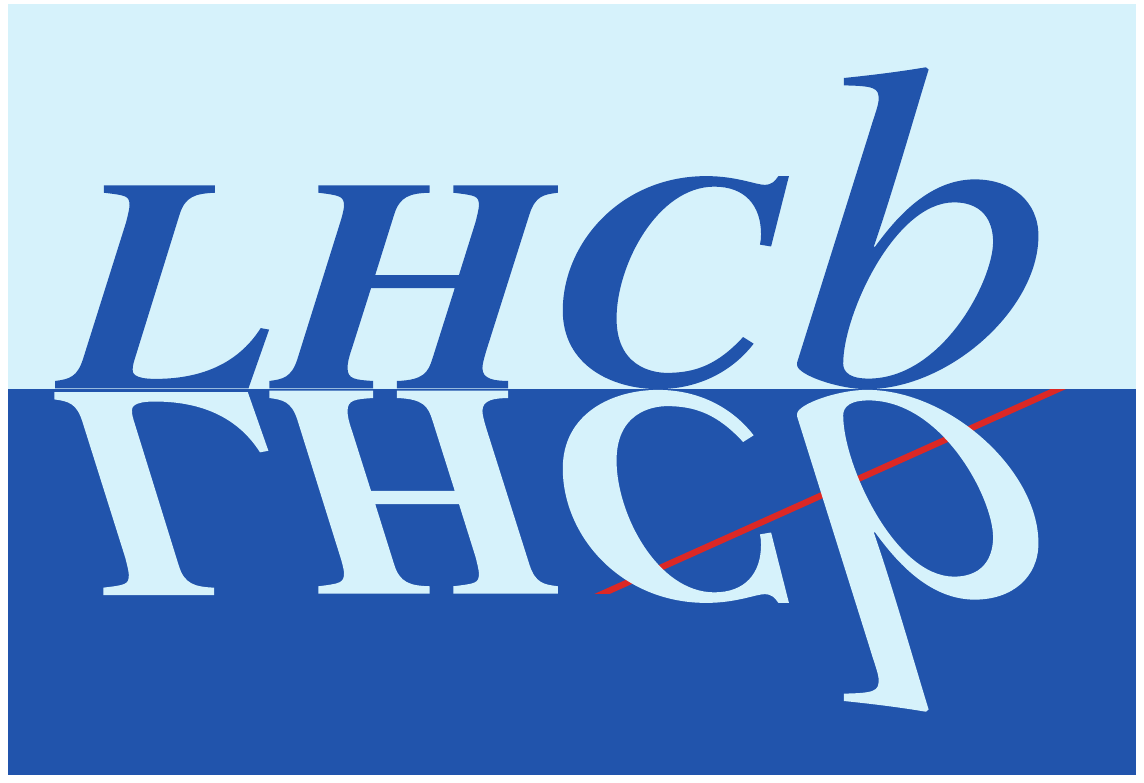}} & &}%
{\vspace*{-1.2cm}\mbox{\!\!\!\includegraphics[width=.12\textwidth]{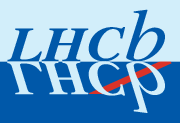}} & &}%
\\
 & & CERN-EP-2016-283 \\  
& & LHCb-PAPER-2016-047 \\  
 & & \today \\ 
& & \\
\end{tabular*}

\vspace*{1.5cm}

{\normalfont\bfseries\boldmath\huge
\begin{center}
  Search for massive long-lived particles decaying semileptonically
  in the LHCb detector
\end{center}
}

\vspace*{1.cm}

\begin{center}
The LHCb collaboration\footnote{Authors are listed at the end of this paper.}
\end{center}

\vspace{\fill}

\begin{abstract}
  \noindent
  A search is presented for massive long-lived particles decaying into a muon and two quarks.
  The dataset consists of proton-proton interactions at
  centre-of-mass energies of 7 and 8\tev, 
  corresponding to integrated luminosities of 1 and 2\invfb, respectively. 
  The analysis is performed assuming a set of production mechanisms with
  simple topologies, including the production of a Higgs-like particle decaying into
  two long-lived particles.
  The mass range from 20 to 80\gevcc and lifetimes from 5 to 100\ps
  are explored.
  Results are also  interpreted in terms of neutralino production in different
  R-Parity violating supersymmetric models,
  with masses in the 23--198~GeV/$c^2$ range.
  No excess above the background expectation is observed and upper limits
  are set on the production cross-section for various
  points in the parameter space of theoretical models.
\end{abstract}

\vspace*{2.0cm}

\begin{center}
  Published on Eur.~Phys.~J.~C (2017) 77:224
\end{center}

\vspace{\fill}

{\footnotesize 
\centerline{\copyright~CERN on behalf of the \lhcb collaboration, licence \href{http://creativecommons.org/licenses/by/4.0/}{CC-BY-4.0}.}}
\vspace*{2mm}

\end{titlepage}


\newpage
\setcounter{page}{2}
\mbox{~}

\cleardoublepage


\renewcommand{\thefootnote}{\arabic{footnote}}
\setcounter{footnote}{0}



\setcounter{page}{1}
\pagenumbering{arabic}


%

\newcommand{\mycomment}[1]{}
\newcommand{\sg}{\tilde{\mathrm{g}}}
\newcommand{\sq}{\tilde{\mathrm{q}}}
\newcommand{\sG}{\tilde{\mathrm G}}
\newcommand{\chino}{\tilde{\chi}}
\newcommand{\chio}{\chino^{0}}
\newcommand{\khi}{\ensuremath{\chio_{1}}\xspace}
\newcommand{\chioi}{\chio_{i}}
\newcommand{\chipm}{\chino^{\pm}}
\newcommand{\chip}{\chino^{+}}
\newcommand{\chim}{\chino^{-}}
\newcommand{\chipmi}{\chino^\pm_i}

\newcommand{\h}{\mathrm{h}}
\newcommand{\Ho}{\H^0}
\newcommand{\ho}{\h^0}

\newcommand{\LLP}{\rm{LLP}}
\newcommand{\LLPs}{\rm{LLPs}}

\newcommand{\msquark}{\ensuremath{m_{\tilde{\rm q}}}\xspace}
\newcommand{\msglue}{\ensuremath{m_{\tilde{\rm g}}}\xspace}

\newcommand{\mx}{\ensuremath{m_{\rm ``\tilde{g}"}}\xspace}

\newcommand{\mchi}{\ensuremath{m_{\khi}}\xspace}
\newcommand{\tauchi}{\ensuremath{\tau_{\khi}}\xspace}

\newcommand{\RXY}{\ensuremath{R_{\rm xy}}\xspace}
\newcommand{\ISOL}{\ensuremath{Isolation}\xspace}
\newcommand{\IP}{\ensuremath{d_{\rm IP}}\xspace}

\newcommand{\Pra}{\ensuremath{P\!A}\xspace}
\newcommand{\Prb}{\ensuremath{P\!B}\xspace}
\newcommand{\Prc}{\ensuremath{P\!\,C}\xspace}
\newcommand{\Prd}{\ensuremath{P\!D}\xspace}
\newcommand{\yone}{\ensuremath{\rm 7\,TeV}\xspace}
\newcommand{\ytwo}{\ensuremath{\rm 8\,TeV}\xspace}
\newcommand{\sideband}{background\xspace}
\newcommand{\ntrak}{\ensuremath{N_{\rm track}}\xspace}
\newcommand{\mllp}{\ensuremath{m_{\LLP}}\xspace}
\newcommand{\taullp}{\ensuremath{\tau_{\LLP}}\xspace}
\newcommand{\mhzero}{\ensuremath{m_{\rm h^0}}\xspace}
\newcommand{\sigr}{\ensuremath{\sigma_{\rm R}}\xspace}
\newcommand{\sigz}{\ensuremath{\sigma_{\rm Z}}\xspace}
\newcommand{\mrec}{\ensuremath{m_{\LLP}^{\rm rec}}\xspace}
\newcommand{\ntrakmin}{\ensuremath{N^{\rm track}_{\rm min}}\xspace}
\newcommand{\mllpmin}{\ensuremath{m^{\rm LLP}_{\rm min}}\xspace}
\newcommand{\sigrmax}{\ensuremath{\sigma^{\rm R}_{\rm max}}\xspace}
\newcommand{\sigzmax}{\ensuremath{\sigma^{\rm Z}_{\rm max}}\xspace}

\newcommand{\plhcb}{{\makebox[1.1\width]{LHCb}}}
\newcommand{\plhcbs}{ \scalebox{0.9}{{\makebox[1.1\width]{LHCb}}} }
\newcommand{\sos}{ \scalebox{0.8}{{\makebox[1.1\width]{\sqs=8 TeV}}} }

\newcommand{\kbb}{\khi\rightarrow\nu\b\bbar}
\newcommand{\kmm}{\khi\rightarrow\nu\mu^+\mu^-}
\newcommand{\kmq}{\khi\rightarrow\mu^\pm\q\q'}
\newcommand{\higgskhi}{\ho\rightarrow\khi\khi}
\newcommand{\higgskhijets}{\ho\rightarrow\khi\khi\rightarrow 6\ \mbox{jets}}
\newcommand{\khijets}{\khi\rightarrow 3\ \mbox{jets}}
\newcommand{\Zkhijets}{\Zo\rightarrow\khi\khi\rightarrow 6\ \mbox{jets}}
\newcommand{\Zkhi}{\Zo\rightarrow\khi\khi}
\newcommand{\hssbbbb}{\ho\rightarrow\s\s\rightarrow\b\bbar\b\bbar}

\def\LSP{LSP}                             
\def\tanb{$\tan{\beta}$}                  
\def\mzero{$\m_0$}                        
\def\mhalf{$\m_{1/2}$}                     
\def\mazero{$\m_{A_0}$}                    
\def\sgnmu{sign($\mu$)}                   

\section{Introduction}
Supersymmetry (SUSY) is one of the most popular extensions of the Standard Model,
which solves the hierarchy problem, can unify the gauge couplings and could provide dark matter candidates.
The minimal supersymmetric standard model (MSSM) is the simplest phenomenologically viable
realization of SUSY~\cite{MSSM,Susy-Martin}.
The present study focuses on a subset of models featuring massive
long-lived particles (LLP) with a measurable flight distance~\cite{KaplanDisplaced2012,hv2}.
LLP searches have been performed by Tevatron and LHC
  experiments~\cite{D0-HV, CDF-HV, ATLAS-HV, art:atlassemi, CMS-dijet,
LHCb-PAPER-2014-062,LHCb-PAPER-2016-014},
often using the Hidden Valley framework~\cite{hv2} as a benchmark model
(see also the study of Ref.~\cite{art:graham}).
The LHCb detector probes the forward rapidity region
which is only partially covered by the other LHC experiments, and
triggers on particles with low transverse momenta, which
allows the experiment to explore relatively small LLP masses.

In this paper a search for massive
long-lived particles is presented, using proton-proton collision data collected by the LHCb
detector at $\sqs=7$ and 8\tev, corresponding to integrated
luminosities of 1 and 2~\invfb, respectively.
The event topology considered in this study is a displaced vertex with several
tracks including a high \pt muon.
This topology is found in the context of the minimal super-gravity (mSUGRA)
realisation of the MSSM,
with R-parity violation~\cite{art:msugra2}, in which the neutralino can decay
into a muon and two jets.
Neutralinos can be produced by a variety of processes.
In this paper four simple production mechanisms with representative
topologies and kinematics are considered, with
the assumed LLP mass in the range 20--80\gevcc.
The LLP lifetime range considered is 5--100\ps, \ie larger than the typical $b$-hadron lifetime.
It corresponds to an average flight distance of up to 30\cm, well inside the \lhcb vertex detector.
One of the production mechanisms considered in detail
is the decay into two LLPs of a Higgs-like particle with an
assumed mass between 50 and 130\gevcc,
\ie in a range which includes the mass of the scalar boson discovered by the ATLAS and CMS
experiments~\cite{Aad:2012tfa,Chatrchyan:2012ufa}.
In addition, inclusive analyses are performed assuming the full set of neutralino
production mechanisms available in \pythia~6\cite{Sjostrand:2006za}.
In this case the LLP mass explored is in the range 23--198\gevcc, inspired by Ref.~\cite{art:graham},
and different combinations of gluino and squark masses are studied.

\section{Detector description}\label{sec:Detector}
The \lhcb detector~\cite{Alves:2008zz,LHCb-DP-2014-002} is a single-arm forward
spectrometer covering the \mbox{pseudorapidity} range $2<\eta <5$,
designed for the study of particles containing \bquark or \cquark
quarks.
The detector includes a high-precision tracking system
consisting of a silicon-strip vertex detector surrounding the $pp$
interaction region (\velo), 
a large-area silicon-strip detector located
upstream of a dipole magnet with a bending power of about
$4{\mathrm{\,Tm}}$, and three stations of silicon-strip detectors and straw
drift tubes, placed downstream of the magnet.
The tracking system provides a measurement of momentum, \ptot, of charged particles with
a relative uncertainty that varies from 0.5\% at low momentum to 1.0\% at 200\gevc.
The minimum distance of a track to a primary vertex (PV), the impact parameter \IP,
is measured with a resolution of $(15+29/\pt)\mum$,
where \pt is the component of the momentum transverse to the beam axis, in\,\gevc.
Different types of charged hadrons are distinguished using information
from two ring-imaging Cherenkov detectors.
Photons, electrons and hadrons are identified by a calorimeter system consisting of
scintillating-pad and preshower detectors, an electromagnetic
calorimeter and a hadronic calorimeter. Muons are identified by a
system composed of alternating layers of iron and multiwire
proportional chambers.
The online event selection is performed by a trigger~\cite{LHCb-DP-2012-004}, 
which consists of a hardware stage based on information from the calorimeter and muon
systems, followed by a software stage which runs a simplified version of the offline event reconstruction.

\section{Event generation and detector simulation}\label{sec:evtgen}

Several sets of simulated events are used to design and optimize the signal
selection and to estimate the detection efficiency.
Proton-proton collisions are generated in \pythia~6
with a specific \lhcb configuration~\cite{LHCb-PROC-2010-056},
and with parton density functions taken from CTEQ6L~\cite{cteq6l}.
The LLP signal in this framework is represented by the lightest neutralino \khi,
with mass \mllp and lifetime \taullp.
It is allowed to decay into two quarks and a muon.
Decays to all quark pairs are assumed to have identical branching 
fractions except for those involving a top quark, which are neglected.

Two separate detector simulations are used to produce signal models:
a full simulation, where the interaction of the generated particles with the detector
is based on \geant~\cite{Allison:2006ve, *Agostinelli:2002hh},
and a fast simulation.
In \geant, the detector and its response are implemented as described in Ref.~\cite{LHCb-PROC-2011-006}.
In the fast simulation, which is used to cover a broader parameter space of the theoretical models,
the charged particles falling into the geometrical acceptance of the detector
are processed by the vertex reconstruction algorithm.
The simulation accounts for the effects of the material veto described in the next section.
The program also provides parameterised particle momenta resolutions, but it is found that these
resolutions have no significant impact on the LLP mass reconstruction, nor on the signal detection efficiency.
%
The fast simulation is validated by comparison with the full simulation.
The distributions for mass, momentum and transverse momentum of the reconstructed LLP
and for the reconstructed decay vertex position are in excellent agreement,
as well as the muon momentum and its impact parameter to the PV.
The detection efficiencies predicted by the full and the fast simulation differ by less than 5\%.

\begin{figure}[!t]
\centering
\mbox{
  \centering
  \subfigure{\includegraphics[width=0.2\textwidth]{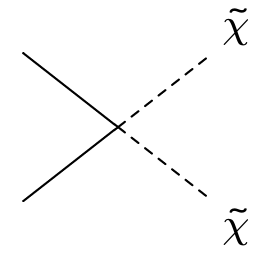} \put(-60,80){\Pra}}
  \subfigure{\includegraphics[width=0.2\textwidth]{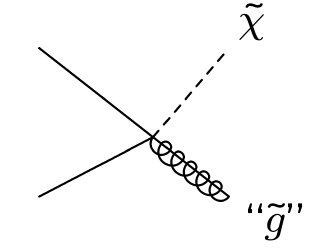}  \put(-60,80){\Prb}}
}
\mbox{
  \centering
  \subfigure{\includegraphics[width=0.2\textwidth]{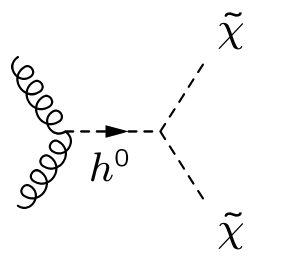}  \put(-60,80){\Prc}}
  \subfigure{\includegraphics[width=0.2\textwidth]{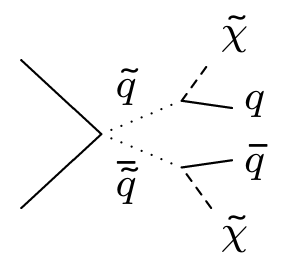}  \put(-60,80){\Prd}}
}
\caption{\small{
    Four topologies considered as representative LLP production mechanisms:
    \Pra non-resonant direct double LLP production,
    \Prb single LLP production,
    \Prc double LLP production from the decay of a Higgs-like boson,
    \Prd double LLP indirect production via squarks.
  }
}
\label{fig:evttopo}
\end{figure}

Two LLP production scenarios are considered.
In the first, the signal samples are generated assuming the full set of neutralino production
processes available in  \pythia.
In particular, nine models are fully simulated with the parameters given in the Appendix,
Table~\ref{tab:models-full}.
Other points in the parameter space of the theoretical models
are studied with the fast simulation, covering the \mllp range 23--198\gevcc.
These models are referred to as ``LV'' (for lepton number violation) followed
by the LLP mass in \gevcc and lifetime (\eg LV98 10\ps).
For the second scenario, the four production mechanisms
depicted in  Fig.~\ref{fig:evttopo}, labelled \Pra, \Prb, \Prc, and \Prd,
are selected and studied independently with the fast simulation.
The LLP, represented by the neutralino, subsequently decays into two quarks and a muon.
The processes \Pra, \Prc, and \Prd have two LLPs in the final state.
In processes \Prc and \Prd two LLPs are produced by the decay of a Higgs-like particle of mass \mhzero, 
and by the decay of squarks of mass \msquark, respectively.
In process \Prb a single LLP is produced recoiling against an object labelled as a ``gluino'', of mass \mx.
In order to control the kinematic conditions, the particles generated in these processes
are constrained to be on-shell and the ``gluino'' of option \Prb is stable.
Since LHCb is most sensitive to relatively low LLP masses, only \mllp values below 80\gevcc are considered.

The background from direct production of heavy quarks,
as well as from $\W$ and $\Z$ boson decays,
is studied using the full simulation.
A sample of $9 \times 10^6$ inclusive \ccbar events with at least two
\cquark hadrons in $1.5<\eta<5.0$, and another sample of about $5 \times 10^5$  \ttbar events
with at least one muon in $1.5<\eta<5.0$ and $\pt>10\gevc$ were produced.
Several million simulated events are available with production of $\W$ and $\Z$ bosons.
The most relevant background in this analysis is from \bbbar events.
The available simulated inclusive \bbbar events are not numerous enough
to cover the high-\pt muon kinematic region required in this
analysis. To enhance the \bbbar background statistics,
a dedicated sample of $2.14 \times 10^5$ simulated events has been produced with a 
minimum parton  $\hat{p}_{\rm T}$ of 20\gevc and requiring a muon with $\pt>12\gevc$ in $1.5<\eta<5.0$.
As a consequence of limitations in the available computing power,
only \bbbar events with $\sqs=7\tev$ have been fully simulated. 
Despite the considerable increase of generation efficiency, all
the simulated \bbbar events are rejected by the multivariate analysis
presented in the next section.
Therefore a data-driven approach is employed for the final background estimation.

\section{Event selection}\label{sec:evtsel}
Signal events are selected by requiring a displaced high-multiplicity vertex
with one associated isolated high-\pt muon, since, due to the larger particle mass,
muons from LLP decays are expected to have larger transverse momenta and to be
more isolated than muons from hadron decays.

The events from $pp$ collisions are selected online by a trigger
requiring muons with $\pt>10\gevc$.
Primary vertices and displaced vertices are reconstructed offline
from charged particle tracks~\cite{LHCb-PUB-2014-044} with a minimum reconstructed \pt of 100\mevc.
Genuine PVs are identified by a small radial distance from the beam axis, $\RXY<0.3$~mm.
The offline analysis requires that the triggering muon has an impact parameter to all PVs
of $\IP>0.25\mm$ and $\pt >12$\gevc.
To suppress the background due to kaons or pions punching through the calorimeters and being misidentified as muons,
the corresponding energy deposit in the calorimeters must be less than 4\% of the muon energy.
To preserve enough background events in the signal-free region for
the signal determination algorithm described in Sect.~\ref{sec:signal-extraction}, no
isolation requirement is applied at this stage.
Secondary vertices are selected by requiring $\RXY>0.55$~mm, at least four tracks in the
forward direction (\ie in the direction of the spectrometer) including the muon and no tracks in the backward
direction.
The total invariant mass of the tracks coming from a selected vertex must be larger than 4.5\gevcc.
Particles interacting with the detector material are an important source of background.
A geometric veto is used to reject events with vertices in regions
occupied by detector material~\cite{LHCb-PAPER-2012-023}.

\begin{figure}[]
  \centering
  \vspace{-5mm}
\includegraphics[clip, trim=0.5mm 1.0mm 0.5mm 0.0mm,height=4.53cm]{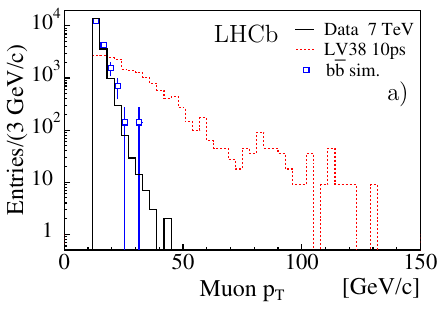}
\includegraphics[clip, trim=0.5mm 1mm 0.5mm 0.0mm,height=4.45cm]{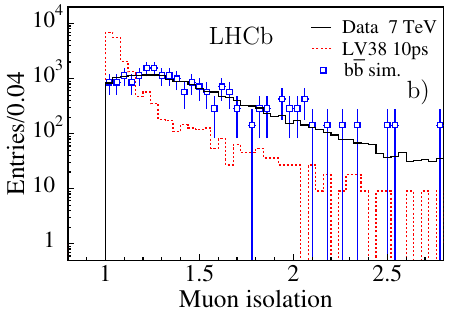}
\includegraphics[clip, trim=0.mm 1.0mm 1mm 0.0mm,height=4.47cm]{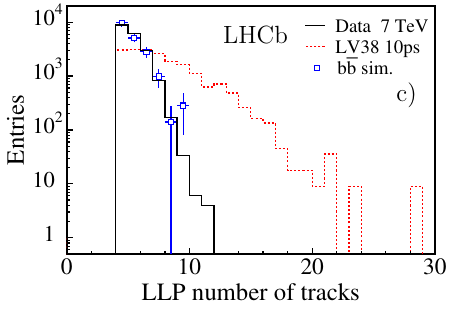}
\includegraphics[clip, trim=0.mm 0.5mm 1mm 0.5mm,height=4.5cm]{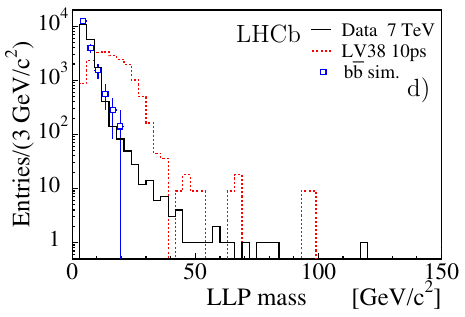}
\includegraphics[clip, trim=0.mm 0.5mm 1mm 0.5mm,height=4.5cm]{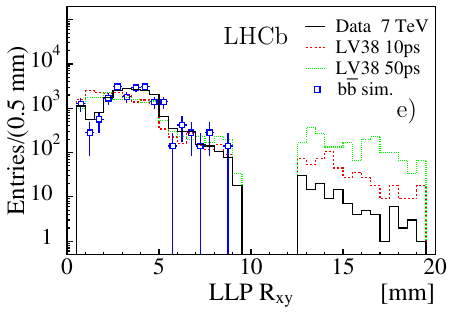}\\
\includegraphics[clip, trim=0.mm 0.5mm 1mm 0.5mm,height=4.5cm]{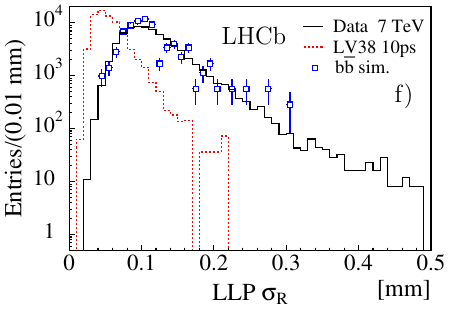}
\includegraphics[clip, trim=0.mm 0.5mm 1mm 0.5mm,height=4.5cm]{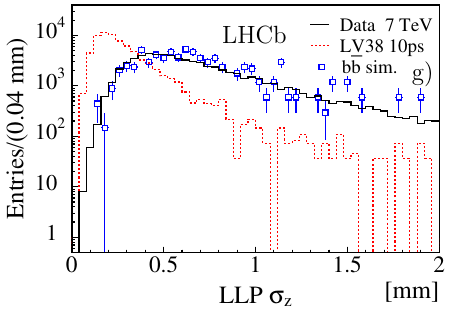}
\caption{ \small
  Distributions for the 7~TeV dataset (black histogram) compared to simulated \bbbar events (blue
  squares with error bars), showing
a) transverse momentum and b) isolation of the muon,
c) number of tracks of the displaced vertex, d)  reconstructed mass, e) radial position of the vertex,
f) and g) vertex fit uncertainties in the radial and z directions.
The fully simulated signal distributions for LV38 10\ps are shown (red dashed histograms), as well as
LV38 50\ps (green dotted histogram) in e).
The distributions from simulation are normalised to the number of data entries.
}
\label{fig:preselrcut}
\end{figure}

The number of data events selected is 18\,925 (53\,331) in the \yone (\ytwo) datasets.
Less than 1\% of the events have more than one candidate vertex, in which case the candidate
with the highest-\pt muon is chosen.
According to the  simulation, the background is largely dominated by \bbbar events,
while the contribution from the decays of $\W$ and $\Z$ bosons is of the order of 10 events.
All simulated \ccbar and \ttbar events are rejected.
The \bbbar cross-section value measured by LHCb,
$ 288 \pm 4 \pm 48$ $\mu$b~\cite{ LHCb-PAPER-2011-003, LHCb-PAPER-2010-002},
predicts  $(15 \pm 3)\times10^3$ events for the \yone dataset, after selection.
The value for the \ytwo dataset is  $(52 \pm 10)\times10^3$.
The extrapolation of the cross-section from \yone to \ytwo is
obtained from POWHEG~\cite{powheg0,*powheg1,*powheg2},
while \pythia is used to obtain the detection efficiency.
The candidate yields for the two datasets are consistent with a dominant
\bbbar composition of the background.
This is confirmed by the study of the shapes of the distributions of the relevant observables.
Figure~\ref{fig:preselrcut} compares the distributions for the \yone dataset
and for the 135 simulated \bbbar events surviving the selection.
For illustration, the shapes of simulated LV38 10\ps signal events are superimposed on all                                         
the distributions, as well as the expected shape for LV38 50\ps on the \RXY distribution.
The muon isolation variable is defined as the sum of the energy of tracks
surrounding the muon direction, including the muon itself, in a cone of radius $R_{\eta \phi}= 0.3$
in the  pseudorapidity-azimuthal angle $(\eta, \phi)$ space, divided by the energy of the muon track.
The corresponding distribution is shown in Fig.~\ref{fig:preselrcut} b).
A muon isolation value of unity denotes a fully isolated muon.
As expected, the muon from the signal is found
to be more isolated than the hadronic background.
Figure~\ref{fig:preselrcut} e) presents the radial distribution of the displaced vertices;
the drop in the number of candidates with a vertex above $\RXY\sim5\mm$ is due to the material veto.
From simulation, the veto introduces a loss of efficiency of 13\% (42\%)
for the detection of LLPs with a $30\gevcc$ mass and a 10\ps (100\ps) lifetime.
The radial ($\sigma_{\rm R}$) and longitudinal ($\sigma_{\rm z}$, parallel to the beam) uncertainties provided
by the LLP vertex fit are shown in Figs.~\ref{fig:preselrcut} f) and g).
Larger uncertainties are expected from the vertex fits of candidates from \bbbar events
compared to signal LLPs. The former are more boosted and produce more narrowly collimated tracks,
while the relatively heavier signal LLPs decay into more divergent tracks.
This effect decreases when \mllp approaches the mass of \bquark-quark hadrons.

A multivariate analysis based on a multi-layer perceptron (MLP)~\cite{McCulloch1943,Hopfield1982}
is used to further purify the data sample.
The MLP input variables are the muon \pt and impact parameter,
the number of charged particle tracks used to reconstruct the LLP,  the vertex radial distance \RXY from the beam line,
and the uncertainties $\sigma_{\rm R}$ and $\sigma_{\rm z}$ provided by the LLP vertex fit.
The muon isolation value and the reconstructed mass of the long-lived particles are not used
in the MLP classifier; the discrimination power of these two variables is subsequently
exploited for the signal determination.
The signal training and test samples are obtained from simulated signal events selected
under the same conditions as data.
A data-driven approach is used to provide the background training samples,
based on the hypothesis that the amount of signal in the data is small.
For this, a number of candidates equal to the number of candidates of the signal training set,
which is of the order of 1000, is randomly chosen in the data.
The same procedure provides the background test samples.
The MLP training is performed independently for each fully simulated model and dataset.
The optimal MLP requirement is subsequently determined by maximizing a figure of merit defined by
$\epsilon / \sqrt{N_d + 1}$, where $\epsilon$ is
the signal efficiency from simulation for a given selection,
and $N_d$ the corresponding number of candidates found in the data.

The generalisation power of the MLP is assessed 
by verifying that the distributions of the classifier output for the
training sample and the test sample agree.
The uniformity over the dataset is controlled by the comparison of
the MLP responses for several subsets of the data.

The MLP classifier can be biased by the presence of signal in the data events used as
background training set. To quantify the potential bias,
the MLP training is performed adding a fraction of simulated signal events (up to 5\%)
to the background set.
This test, performed independently for all signal models, demonstrates a negligible
variation of the performances quantified by the above figure of merit.

\section{Determination of the signal yield}\label{sec:signal-extraction}
\begin{figure}[!]
  \centering
  \hspace{-2mm}
{\includegraphics[clip, trim=0.5mm 0.5mm 0.5mm 0.5mm, height=4.5cm]{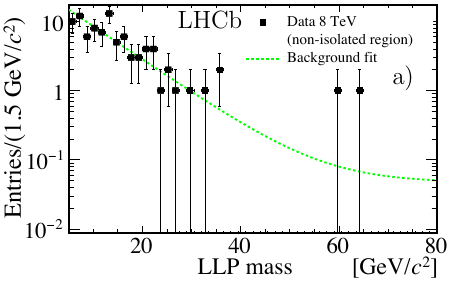}}
{\includegraphics[clip, trim=0.5mm 0.5mm 0.5mm 0.7mm, height=4.5cm]{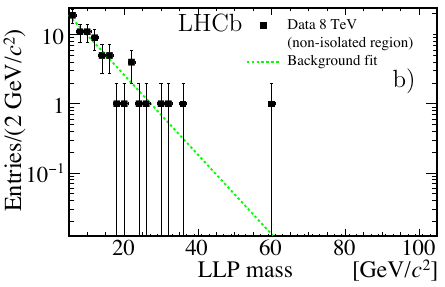}}
{\includegraphics[clip, trim=0.5mm 0.2mm 0.5mm 0.2mm, height=4.5cm]{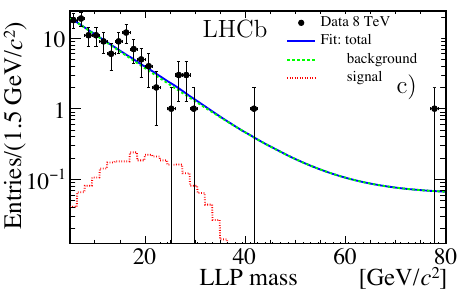}}
{\includegraphics[clip, trim=0.5mm 0.5mm 0.5mm 0.1mm, height=4.5cm ]{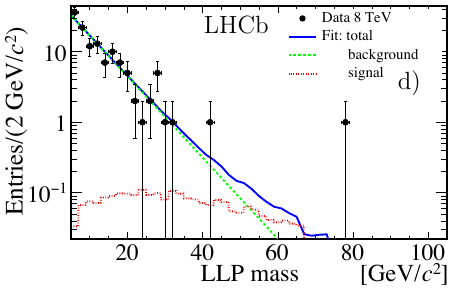}}
\caption{\small
  Reconstructed mass of the LLP candidate from the 8 TeV dataset.
  The top plots correspond to events with candidates selected
  from the \sideband region of the muon isolation variable.
  They are fitted with the sum of two exponential functions.
  In the bottom row the candidates from the signal region are fitted including a specific signal shape,
  added to the background component.
  Subfigures a) and c) correspond to the analysis which assumes the LV38 5\ps signal model,
  b) and d) are for LV98 10\ps.
}
\label{fig:fit_2012}
\end{figure}

The signal yield is determined with an extended unbinned maximum likelihood fit
to the distribution of the reconstructed LLP mass, with
the shape of the signal component taken from the simulated models, plus the background component.
After the MLP filter, no simulated background survives;
therefore a  data-driven method is adopted to determine the background template.
The data candidates are separated into a signal region with muon isolation below 1.4
and a \sideband region with isolation value from 1.4 to 2.
The signal region contains more than 90\% of the signal for all the models considered
(see \eg Fig.~\ref{fig:preselrcut} (b)).
The reconstructed mass obtained from the \sideband candidates is used to constrain
an empirical  probability density function (PDF) consisting of the sum of
two negative slope exponential functions,
for which the slope values and amplitudes are free parameters in the fit. 
The signal PDF  is taken from the histogram of the mass distribution
obtained from simulation.
The fit is performed simultaneously on events from the \sideband region
and from the signal region. In the latter the numbers of signal and background
events are left free in the fit, while the slope values and the relative strength of the two
exponential functions are in common with the \sideband region fit.
Examples of fit results are given in Fig.~\ref{fig:fit_2012}, obtained from the \ytwo dataset
for two signal hypotheses,  LV38 5\ps and LV98 10\ps.
The fitted signal yields, given in Table~\ref{tab:fitH-res}, for both datasets are compatible
with the background-only hypothesis.

\begin{table}
  \caption{\small
    Total signal detection efficiency $\epsilon$, including the geometrical acceptance,
    and numbers of fitted signal and background events, $N_{\rm s}$ and $N_{\rm b}$,
    for the different  signal hypotheses.
    The last column gives the value of   \chisqndf  from the fit.
    The signal models are from the full simulation.
    Uncertainties are explained in Sect.~\ref{sec:eff}.
   }
 \label{tab:fitH-res} 
 \centering
 {\small  
\begin{tabular}{clccccc}
\hline
Dataset & Model     &  $\epsilon $ \% & $N_{\rm b}$ & $N_{\rm s}$&  \chisqndf \\
\hline
\yone
&LV 38  5\ps & $    0.52 \pm   0.03 $  & $    140.2 \pm   15.5 $  & $ \phantom{-}    3.8 \pm   10.0 $  &  0.64 \\
&LV 38 10\ps & $    0.57 \pm   0.03 $  & $    115.2 \pm   13.3 $  & $ \phantom{-}    4.8 \pm\phantom{0}    8.2 $  &  1.71 \\
&LV 38 50\ps & $    0.43 \pm   0.02 $  & $    112.9 \pm   13.3 $  & $ \phantom{-}    9.0 \pm \phantom{0}   8.6 $  &  1.50 \\
&LV 98  5\ps & $    0.58 \pm   0.03 $  & $ \phantom{0}97.3 \pm 10.3 $            & $ -3.3 \pm \phantom{0}  2.4 $  &  0.88 \\
&LV 98 10\ps & $    0.72 \pm   0.04 $  & $ \phantom{0}62.6 \pm \phantom{0}8.7 $  & $ -5.6 \pm \phantom{0}  2.8 $  &  1.06 \\
&LV 98 50\ps & $    0.56 \pm   0.03 $  & $ \phantom{0}99.9 \pm   11.2 $          & $ -3.9 \pm \phantom{0}  4.5 $  &  0.33 \\
&LV198  5\ps & $    0.60 \pm   0.04 $  & $    143.8 \pm   12.5 $  & $               -6.9 \pm  \phantom{0}  2.2 $  &  1.42 \\
&LV198 10\ps & $    0.76 \pm   0.04 $  & $    158.1 \pm   13.1 $  & $               -6.1 \pm  \phantom{0}  2.7 $  &  1.63 \\
&LV198 50\ps & $    0.66 \pm   0.04 $  & $    118.8 \pm   11.3 $  & $               -0.9 \pm \phantom{0}   2.8 $  &  0.89 \\
\hline
\ytwo
&LV 38  5\ps & $    0.54 \pm   0.04 $  & $    120.3 \pm   15.6 $  & $ \phantom{-}    2.9 \pm \phantom{0}   9.5 $  &  0.74 \\
&LV 38 10\ps & $    0.66 \pm   0.04 $  & $    203.7 \pm   19.9 $  & $               -1.6 \pm   13.3 $  &  0.81 \\
&LV 38 50\ps & $    0.43 \pm   0.02 $  & $    123.3 \pm   15.6 $  & $ \phantom{-}    3.7 \pm   11.0 $  &  0.99 \\
&LV 98  5\ps & $    0.77 \pm   0.05 $  & $    121.0 \pm   11.2 $  & $ \phantom{-}    1.0 \pm \phantom{0}   2.2 $  &  1.26 \\
&LV 98 10\ps & $    0.96 \pm   0.05 $  & $    123.7 \pm   12.0 $  & $ \phantom{-}    2.4 \pm \phantom{0}   3.4 $  &  0.74 \\
&LV 98 50\ps & $    0.69 \pm   0.04 $  & $    103.8 \pm   10.5 $  & $ \phantom{-}    2.2 \pm  \phantom{0}  2.8 $  &  0.94 \\
&LV198  5\ps & $    0.79 \pm   0.06 $  & $    196.3 \pm   14.2 $  & $               -2.3 \pm  \phantom{0}  2.0 $  &  1.94 \\
&LV198 10\ps & $    1.06 \pm   0.07 $  & $    258.7 \pm   16.2 $  & $ \phantom{-}    2.3 \pm  \phantom{0}  2.3 $  &  1.53 \\
&LV198 50\ps & $    0.69 \pm   0.04 $  & $    113.7 \pm   10.8 $  & $ \phantom{-}    1.3 \pm  \phantom{0}  2.1 $  &  1.73 \\
\hline
 \end{tabular}
 }
\end{table}

The validity of using events with isolation above 1.4 to model the background has been
checked by comparing the relevant distributions from events in the background and in the signal regions,
including the muon \pt and impact parameter distributions, as well as
the number of tracks, invariant mass, vertex \RXY and vertex uncertainties of the LLP candidate.
This test is performed with the nominal MLP selection,
and also with loosened requirements that result in a threefold increase
in the number of background candidates.
In both cases all distributions agree within statistical uncertainties,
with the \chisqndf of the comparison in the range 0.6-1.5.

The sensitivity of the procedure is studied by adding a small number of signal
events to the data according to a given signal model. The fitted yields are
consistent with the numbers of added events on average, and the pull distributions are
close to Gaussian functions
with mean values between $-0.1$ and $0.1$ and standard deviations in the range from 0.9 to 1.2.

As a final check a two-dimensional sideband subtraction method (``ABCD method''~\cite{ABCD})
has been considered. The LLP reconstructed mass and the muon isolation are used
to separate the candidates in four regions. The results of this check 
are also consistent with zero signal for the two datasets.

Both the LLP mass fit and the ABCD methods are tested with $W$ and $\Z/\gamma$ leptonic decays.
Isolated high-\pt muons are produced in such processes with
kinematic properties similar to the signal.
By removing  the minimum \RXY requirement the candidates can be formed by collecting tracks from the primary vertex.
As before, the \sideband is taken from a region of  muon isolation above 1.4,
which contains a negligible amount of signal.
For both datasets the number of events obtained from this study is compatible with the
cross-sections measured by \lhcb~\cite{LHCb-PAPER-2015-049,LHCb-PAPER-2015-001,LHCb-PAPER-2014-033}.

\section{Detection efficiency and systematic uncertainties}\label{sec:eff}

The total signal detection efficiency, estimated from fully simulated events, is shown
in Table~\ref{tab:fitH-res}. It includes the geometrical acceptance, which
for the detection of one \khi in LHCb is about 11\% (12\%) at $\sqs=7\tev$ (8\tev).
The efficiencies for the models where the fast simulation is used,
including processes \Pra, \Prb, \Prc, and \Prd, vary from
about 0.1\% to about 2\%.
The efficiency increases with \mllp because more particles are produced in the decay of heavier LLPs.
This effect is only partially counteracted by the
loss of particles outside the spectrometer acceptance,
which is especially likely when the LLP are produced from the decay of heavier states,
such as the Higgs-like particles of process \Prc.
Another competing phenomenon is that
the lower boost of heavier LLPs results in a shorter average flight length, \ie
the requirement of a minimum \RXY disfavours heavy LLPs.
The cut on \RXY is more efficient at selecting LLPs with large lifetimes, but 
for lifetimes larger than $\sim50\ps$
a considerable portion of the decays falls into the material region and is vetoed.
Finally, a drop of sensitivity is expected for LLPs with a
lifetime close to the \bquark hadron lifetimes,
where the contamination from \bbbar events becomes important, especially for low mass LLPs.


A breakdown of the relative systematic uncertainties for the analysis of the \ytwo
dataset is shown in Table~\ref{tab:sys-small}.
The table does not account for the uncertainties associated with the fit procedure,
which, as described below, require a specific treatment.
The uncertainties on the integrated luminosity are $1.7\%$ for \yone dataset and $1.2\%$ for \ytwo data~\cite{art:lumi}.  Several sources of systematic uncertainty coming from discrepancies between data and simulation have been considered.

The muon detection efficiency, including trigger, tracking, and muon identification efficiencies,
is studied by a tag-and-probe technique applied to muons
from \mbox{$J/\psi \rightarrow \mumu$~\cite{LHCb-DP-2013-002}}  and
from $\Z \rightarrow \mumu$ decays~\cite{LHCb-PAPER-2012-008,LHCb-PAPER-2014-033,LHCb-PAPER-2015-001,LHCb-PAPER-2015-049}.
The corresponding systematic effects
due to differences between data and simulation are estimated to be between 2.1\% and 4.5\%,
depending on the theoretical model considered.

A comparison of the simulated and observed \pt distributions of muons from \mbox{$\Z \rightarrow \mumu$} decays
shows a maximum difference of 3\% in the momentum scale;
this difference is propagated to the LLP analysis by moving the muon \pt threshold by the same amount.
A corresponding systematic uncertainty of 1.5\% is estimated for all models under consideration.

The \IP distribution shows a discrepancy between data and simulation of about 5\mum
in the mean value for muons from $\Z$ decays,
with a maximum deviation of about 20\mum close to the muon \pt threshold.
By changing the minimum \IP requirement by this amount,
the change in the detection efficiency is in the range 0.4--1.2\%, depending on the model.

The vertex reconstruction efficiency is affected by the tracking efficiency and
has a complicated spatial structure due to the geometry of the \velo and the material veto.
In the material-free region, $\RXY<4.5\mm$,
the efficiency to detect secondary vertices as a function of the flight distance
has been studied in detail, in particular in the context of the
\bquark hadrons lifetime measurement~\cite{LHCb-PAPER-2013-065}. The deviation of
the efficiency in simulation with respect to the data is below 1\%. 
For \RXY from 4.5\mm to about 12\mm a study performed with inclusive \bbbar events
finds differences between data and simulation of less than 5\%.
The corresponding systematic uncertainties are determined by
altering the efficiency in the simulation program as a function of the true vertex position.
A maximum of 1\% uncertainty is obtained for all the signal models.
An alternative procedure to asses this uncertainty considers vertices from
$B^0 \rightarrow \jpsi \Kstarz$ decays with $\decay{\jpsi}{\mumu}$ and $\Kstarz \rightarrow K^+ \pi^-$.
The detection efficiency in data and simulation is found to agree within 10\%.
This result, obtained from a four-particle final state, when propagated
to LLP decays with on average more than 10 charged final-state particles
for all modes, results in a discrepancy of at most 2\% between the LLP
efficiencies in data and simulation, which is the adopted value for the
respective systematic uncertainty.

The uncertainty on the position of the beam line is less than 20\mum~\cite{LHCb-DP-2014-001}.
It can affect the secondary vertex selection, mainly via the requirement on \RXY.
By altering the PV position in simulated signal events,
the maximum effect on the LLP selection efficiency is in the range 0.2--1\%.

The imprecision of the models used for training the MLP propagates into a systematic difference of
the detection efficiency between data and simulation.
The bias on each input variable is determined by comparing simulated and experimental
distributions for muons and LLP candidates from  \Z and \W events, and 
from \bbbar events.
The effect of the biases is subsequently estimated by testing the trained classifier
on altered simulated signal events:
each input variable is modified by a scale factor randomly drawn from a
Gaussian distribution of width equal to the corresponding bias.
The RMS variation of the signal efficiency distributions after the MLP range
from 1.5 to 3.6\% depending on the signal model.
These values are taken as contributions to the systematic uncertainties.

The signal region is selected by the requirement of a muon isolation value
lower than 1.4. By a comparison of data and simulated muons from \Z decays,
the uncertainty on this variable is estimated to be $\pm 0.05$.  
This uncertainty is propagated to a maximum 2.2\% effect on the detection efficiency.

Comparing the mass  distributions of \bbbar events selected with relaxed cuts, a maximum
mass scale discrepancy between data and simulated events of 10\% is estimated.
The corresponding shift of the simulated signal mass distribution results
in a variation of the detection efficiency between 0.8 and 1.5\%.

The statistical precision of the efficiency value determined from the simulated events
is in the range 1.7--2.5\% for the different models.
 
The theoretical uncertainties are dominated by the uncertainty of the partonic luminosity.
Their contribution to the detection efficiency uncertainty is
estimated following the procedure explained in Ref.~\cite{PDF4LHC} and vary
from 3\% up to a maximum of 7\%,  which is found for the gluon-gluon fusion process \Prc.

For the analysis based on the fast simulation, a 5\% uncertainty is added to account 
for the difference between the fast and the full simulation, as explained in Sect.~\ref{sec:evtgen}.

\begin{table}[t]
\begin{center}
\caption{\small
  Summary of the contributions to the relative systematic uncertainties,  corresponding to the 8 TeV dataset,
  (the sub-total for the 7 TeV dataset is also given).
  The indicated ranges cover the fully simulated LV models.
  The detection efficiency is affected by the parton luminosity model and depends upon the production process, with a maximum
  uncertainty of 7\% for the gluon-gluon fusion process \Prc.
  For the fast simulation based analysis there is an additional contribution of 5\%.
  The systematic effects associated with the signal and background models used in the
  LLP mass fit are not shown in the table.
}
\scalebox{0.9}{
\begin{tabular}{lc}
\hline
Source & Contribution (\%)\\ 
\hline
Integrated luminosity              &    1.2 \\
Muon detection                     & $   2.1 -   4.5$ \\
Muon \pt scale                     & $   1.5 $ \\
Muon \IP uncertainty                & $   0.4 -   1.2$ \\
Vertex reconstruction              &    2.0 \\
Beam line uncertainty              & $   0.2 -   1.0$ \\
MLP training models                & $   1.5 -   3.6$ \\
Muon isolation                     & $       2.2$ \\
LLP mass scale                     & $   0.8 -   1.5$ \\
Models statistics                  & $   1.7 -   2.5$ \\
\hline
Sub-total \ytwo dataset            & $   4.9-   6.5$ \\
(Sub-total \yone dataset            & $   4.9-   6.1$) \\
\hline
Parton luminosity                  & $ 3-7 $ \\
Analysis with fast simulation      & $ 5   $ \\
\hline
\end{tabular}
}
\label{tab:sys-small}
\end{center}
\end{table}

  The choice of the background and signal templates can affect the results of the LLP mass fit.
  The uncertainty due to the signal model accounts for the
  mass scale, the mass resolution and the finite number of events available to construct the model.
  Pseudoexperiments in which 10 signal events are added to the data are analysed with a
  modified signal template, and the resulting number of fitted candidates is compared to the
  result from the nominal fit model.
  Assuming as before a 10\% uncertainty on the signal mass scale, a maximum absolute variation of 0.6
  fitted signal candidates is obtained.
  No significant effects are obtained by modifying the signal mass resolution with an additional smearing.
  Changing the statistical precision by reducing 
  the initial number $N$ of signal events used to build the histogram PDF by $2 \sqrt{N}$
  has no significant effect either.

  The uncertainty induced by the choice of the background model is obtained by reweighting
  the candidates from the \sideband region in such a way that the distribution  of the number of tracks
  included in the LLP vertex fit exactly matches the distribution in the signal region.
  This test is motivated by the fact that the number of tracks has a significant
  correlation with the measured mass.
  The fits of the mass distribution of pseudoexperiments give absolute variations in the numbers
  of fitted signal events in the range 0.1--1.6, the largest value at low LLP mass.
  Reweighting the candidates in such a way as to match the \pt distributions gives variations
  which are less than 0.5 events for all models.
  Moving the isolation threshold by $\pm 0.1$ leads to variations of the order of 0.01 events. 
  In conclusion, the variation on the number of fitted candidates associated to the choice of the PDF models
  is in the range of 1--2 events.
  The calculation of the cross-section upper limits takes into account this uncertainty as an additional
  nuisance parameter on the fit procedure.

\section{Results}

\begin{figure}
\centering
\includegraphics[height=4.5cm,angle=0]{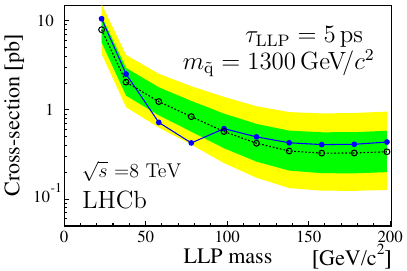}
\includegraphics[height=4.5cm,angle=0]{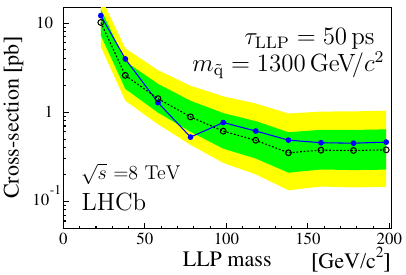}\\
\vspace{3mm}
\includegraphics[height=4.5cm,angle=0]{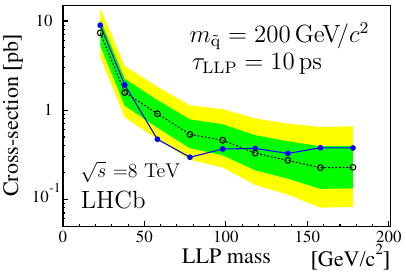}
\includegraphics[height=4.5cm,angle=0]{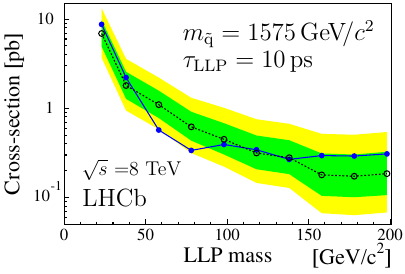}
\caption{ \small
  Expected (open dots with 1$\sigma$ and 2$\sigma$ bands) and observed (full dots)
  cross-section times branching fraction upper limits at 95\% confidence level,
  as a function of the LLP mass from the 8~TeV dataset.
  The theoretical models assume the full set of SUSY production processes available
  in \pythia~6 with default parameter settings, unless otherwise specified.
  The gluino mass is 2000\gevcc.
}
\label{fig:fast_LV}
\end{figure}

\begin{figure}
  \centering
\includegraphics[clip, trim=0.mm 0.mm 0.mm 0.mm, height=4.5cm,angle=0]{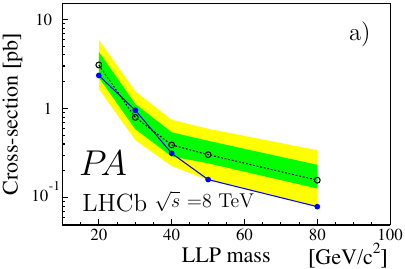}
\includegraphics[clip, trim=0.mm 0.mm 0.mm 0.mm,height=4.5cm,angle=0]{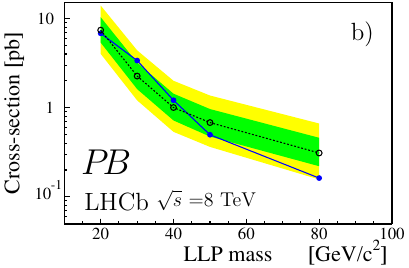}\\
\vspace{3mm}
\includegraphics[clip, trim=0.5mm 0.mm -0.5mm 0.mm,height=4.5cm,angle=0]{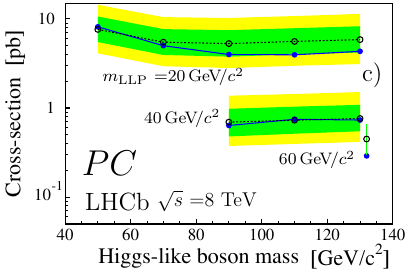}
\includegraphics[clip, trim=0.mm 0.mm 0.mm 0.mm,height=4.5cm,angle=0]{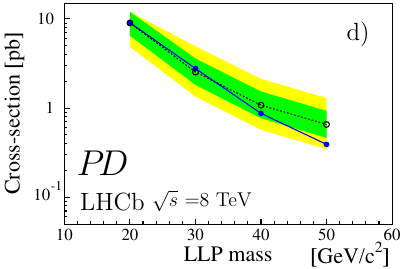}
\caption{ \small
  Expected (open dots and 1$\sigma$ and 2$\sigma$ bands) and observed (full dots)
  cross-section times branching fraction upper limits (95\% CL) for
  the processes indicated in the bottom left corner
  of each plot, \taullp is always 10\ps. The results correspond to the 8~TeV dataset.
  a) upper limits as a function of the LLP mass for process \Pra;
  b) as a function of the LLP mass for process \Prb, with $\mx=100$\gevcc;
  c) as a function of \mhzero for process \Prc for \mllp of 20,
  40, and 60\gevcc, from top to bottom (the single point at 130\gevcc with $\mllp=60\gevcc$
  has been shifted to the right for visualisation);
  d)  upper limits  as a function of the LLP mass for process \Prd with $\msquark=60$\gevcc.
}
\label{fig:topo_ul_1}
\end{figure}

\begin{figure}
\centering
\includegraphics[clip, trim=0.mm 0.mm 0.mm 0.mm, height=4.5cm,angle=0]{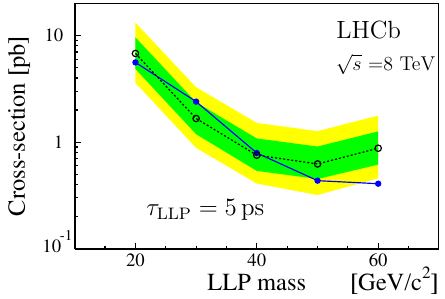}
\includegraphics[clip, trim=0.mm 0.mm 0.mm 0.mm, height=4.5cm,angle=0]{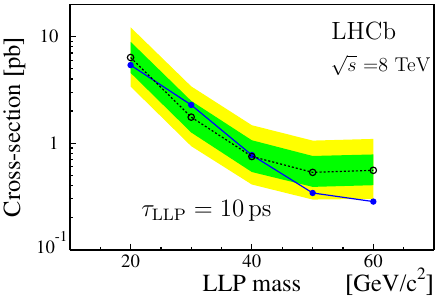}\\
\includegraphics[clip, trim=-0.5mm 0.mm 0.5mm 0.mm, height=4.5cm,angle=0]{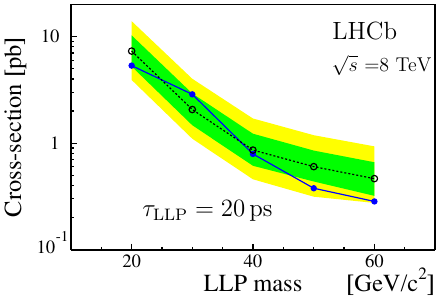}
\includegraphics[clip, trim=0.mm 0.mm 0.mm 0.mm, height=4.5cm,angle=0]{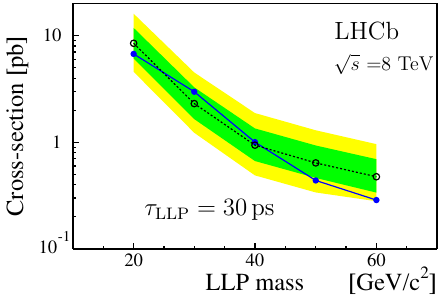}\\
\includegraphics[clip, trim=0.mm 0.mm 0.mm 0.mm, height=4.5cm,angle=0]{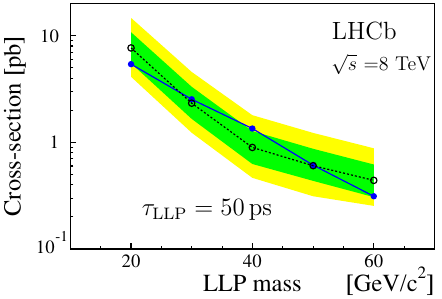}
\includegraphics[clip, trim=0.mm 0.mm 0.mm 0.mm, height=4.5cm,angle=0]{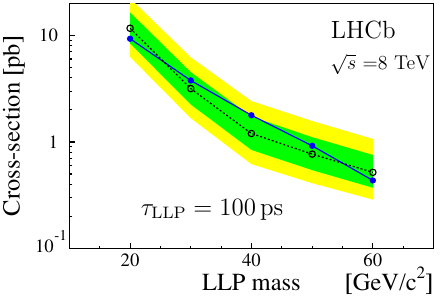}
\caption{ \small
  Expected (open dots with 1$\sigma$ and 2$\sigma$ bands) and observed (full dots)
  cross-section times branching fraction upper limits (95\% CL) for the processes \Prc\ as a function
  of the LLP mass; the LLP lifetime \taullp is indicated in each plot, $\mhzero=125\gevcc.$
  The results correspond to the 8~TeV dataset.
}
\label{fig:topo_h125}
\end{figure}

The LLP candidates collected at $\sqs=7$ and 8\tev are analysed independently.
The fast simulation is used to extend the MSSM/mSUGRA theoretical parameter space of the LV models,
and for the analysis of processes \Pra, \Prb, \Prc, and \Prd.
The results obtained are found to be compatible with the absence of signal for all signal model hypotheses
considered.
The 95\% confidence level (CL) upper limit on the production cross-sections
times branching fraction is computed
for each model using the CLs approach~\cite{art:cls}.
The numerical results for the fully simulated LV models are given in Table~\ref{tab:upperlimit-LV}
\footnote{The numerical results for all the other models are available as supplementary material}.
A graphical representation of selected results is given in Figs.~\ref{fig:fast_LV} to \ref{fig:topo_h125}.

The MSSM/mSUGRA LV models are explored by changing the common squark mass and the gluino mass.
Figure~\ref{fig:fast_LV} gives examples of the cross-section times branching fraction
upper limits as a function of \mllp for such models for two values of \taullp, and two
values of the squark mass. The gluino mass is set to 2000\gevcc. Varying the gluino mass
from 1500 to 2500\gevcc has almost no effect on the results.
The decrease of sensitivity for decreasing \mllp is explained by the above-mentioned effects
on the detection efficiency.

A representation of selected results from the processes \Pra, \Prb, \Prc, and \Prd
is given in Fig.~\ref{fig:topo_ul_1}.
The single LLP production of \Prb has a lower detection probability compared to
the double LLP production case, \Pra, which explains the reduced sensitivity.
The \Prb plots correspond to $\mx=100\gevcc$. Varying \mx from 100 to 1000\gevcc
decreases the detection efficiency by a factor of two, while an increase by a factor of two is
obtained reducing $\mx$ to 20\gevcc.
The results for process \Prc are given as a function of the Higgs-like boson mass, for three values of \mllp.
Again  the sensitivity of the analysis drops with decreasing \mllp.
The results shown for \Prd are for $\msquark=60\gevcc$, which limits the maximum \mllp value.
In process \Prd some of scattering energy is absorbed by an additional jet during the LLP production, 
reducing the detection efficiency by a factor of two with respect to \Pra.
Finally, Fig.~\ref{fig:topo_h125} gives the cross-section upper limits times branching fraction
as a function of \mllp, for the process \Prc with a mass of 125\gevcc for the Higgs-like boson
and LLP lifetime from 5 to 100\ps.
These results can be compared to the prediction of the Standard Model Higgs production cross-section
of about 21\pb at $\sqs=8\tev$~\cite{Heinemeyer:2013tqa}.

\begin{table}
\begin{center}
\caption{  \small
Upper limits (95\% CL) on the production cross-section times branching fraction (pb)
for the \yone and \ytwo datasets, based on the fully simulated LV signal samples.
}
\label{tab:upperlimit-LV}
\scalebox{0.9}{
\renewcommand{\arraystretch}{1.25}
\begin{tabular}{lccccc}
   \hline
&\multicolumn{2}{c}{\yone dataset} && \multicolumn{2}{c}{\ytwo dataset}   \\
\cline{2-3}  \cline{5-6}
  \raisebox{3mm}{Model}          &  Expected      & Observed  &&  Expected      & Observed        \\
  \hline
LV 38  5\ps &     4.03$^{+1.79}_{-1.20}$ &    4.73 &&    2.04$^{+0.89}_{-0.60}$ &    2.32  \\
LV 38 10\ps &     2.95$^{+1.36}_{-0.89}$ &    3.76 &&    2.24$^{+0.95}_{-0.65}$ &    2.13  \\
LV 38 50\ps &     4.08$^{+1.89}_{-1.24}$ &    6.15 &&    2.86$^{+1.23}_{-0.83}$ &    3.10  \\
LV 98  5\ps &     1.78$^{+0.97}_{-0.60}$ &    1.21 &&    0.62$^{+0.36}_{-0.22}$ &    0.57  \\
LV 98 10\ps &     1.52$^{+0.78}_{-0.49}$ &    0.94 &&    0.52$^{+0.27}_{-0.17}$ &    0.53  \\
LV 98 50\ps &     2.21$^{+1.10}_{-0.70}$ &    1.83 &&    0.70$^{+0.41}_{-0.25}$ &    0.77  \\
LV198  5\ps &     1.50$^{+0.86}_{-0.52}$ &    0.95 &&    0.59$^{+0.34}_{-0.21}$ &    0.40  \\
LV198 10\ps &     1.18$^{+0.68}_{-0.41}$ &    0.85 &&    0.27$^{+0.20}_{-0.11}$ &    0.42  \\
LV198 50\ps &     0.92$^{+0.67}_{-0.38}$ &    1.07 &&    0.52$^{+0.35}_{-0.21}$ &    0.58  \\
\hline
\end{tabular}
}
\end{center}
\end{table}

%

\section{Conclusion}
Long-lived massive particles decaying into a muon and two quarks have been searched for
using  proton-proton collision data collected by LHCb at $\sqs =7$ and 8\tev,
corresponding to integrated luminosities of 1 and 2~\invfb, respectively.
The background is dominated by \bbbar events and is reduced by tight selection
requirements, including a dedicated multivariate classifier.
The number of candidates is determined by a fit to the LLP reconstructed mass with a signal shape
inferred from the theoretical models.

LHCb can study the forward region $2<\eta <5$, and its low trigger \pt threshold allows the experiment to
explore relatively small LLP masses.
The analysis has been performed assuming four LLP production mechanisms with the topologies shown
in Fig.~\ref{fig:evttopo}, covering LLP lifetimes  from 5 ps up to 100\ps and masses in the
range  20--80~\gevcc.
One of the processes  proceeds via the decay of a Higgs-like particle into two LLPs:
the mass of the Higgs-like particle is varied between 50\gevcc and 130\gevcc,
comprising the mass of the scalar boson discovered by the ATLAS and CMS experiments.
In addition, the full set of neutralino production mechanisms available in \pythia
in the context of MSSM/mSUGRA has been considered, with an LLP mass range 23--198~\gevcc.
The results for all theoretical models considered are compatible with the background-only hypothesis.
Upper limits at 95\% CL are set on the cross-section times branching fractions.

\clearpage

\clearpage
\section*{Acknowledgements}
 
\noindent We express our gratitude to our colleagues in the CERN
accelerator departments for the excellent performance of the LHC. We
thank the technical and administrative staff at the LHCb
institutes. We acknowledge support from CERN and from the national
agencies: CAPES, CNPq, FAPERJ and FINEP (Brazil); NSFC (China);
CNRS/IN2P3 (France); BMBF, DFG and MPG (Germany); INFN (Italy); 
FOM and NWO (The Netherlands); MNiSW and NCN (Poland); MEN/IFA (Romania); 
MinES and FASO (Russia); MinECo (Spain); SNSF and SER (Switzerland); 
NASU (Ukraine); STFC (United Kingdom); NSF (USA).
We acknowledge the computing resources that are provided by CERN, IN2P3 (France), KIT and DESY (Germany), INFN (Italy), SURF (The Netherlands), PIC (Spain), GridPP (United Kingdom), RRCKI and Yandex LLC (Russia), CSCS (Switzerland), IFIN-HH (Romania), CBPF (Brazil), PL-GRID (Poland) and OSC (USA). We are indebted to the communities behind the multiple open 
source software packages on which we depend.
Individual groups or members have received support from AvH Foundation (Germany),
EPLANET, Marie Sk\l{}odowska-Curie Actions and ERC (European Union), 
Conseil G\'{e}n\'{e}ral de Haute-Savoie, Labex ENIGMASS and OCEVU, 
R\'{e}gion Auvergne (France), RFBR and Yandex LLC (Russia), GVA, XuntaGal and GENCAT (Spain),
Herchel Smith Fund, The Royal Society, Royal Commission for the Exhibition of 1851 and the Leverhulme Trust (United Kingdom).


\clearpage
{\noindent\normalfont\bfseries\Large Appendix}

\appendix
\section*{Parameters of the fully simulated signal models}\label{app:gen}

The parameters used to generate nine fully simulated signal samples in the context
of MSSM/mSUGRA are given in Table~\ref{tab:models-full}.
Other MSSM parameters remain at their default \pythia values.
The lightest neutralino, $\khi$, decays via the lepton number violating mode LQD
(for the definition see~\cite{KaplanDisplaced2012,art:msugra2}).
As an approximation, equal branching fractions are assumed for all QD pairs,
except for the pairs with a top quark, which are excluded.

Two sets of events have been produced with \sqs=7 and 8\tev.
Only events with one muon and one $\khi$ in the LHCb acceptance are processed in \geant,
corresponding to about 11\% of the \sqs=7\tev generated events, 12\% at 8\tev. 

\begin{table}[!h]
  \begin{center}
     \caption{\small{
     Parameters for the generation of the nine fully simulated signal models. 
     The LLP is the lightest neutralino, $\khi$ with  $\mchi = \mllp$;
      $M_1$ and $M_2$ are the \pythia parameters RMSS(1) and RMSS(2), $m_{\tilde{\rm g}}$ is RMSS(3),
     $\mu$ is RMSS(4), \tanb \, RMSS(5) and $m_{\tilde{\rm q}}$ is RMSS(8-12).
     Samples with lifetime of 5, 10 and 50\ps have been produced for each mass.
    }}
     \scalebox{0.95}{
     \begin{tabular}{l c c c c c c c}
      \hline
      Model  & $M_1$  & $M_2$  & \msglue  & \tanb &  $\mu$   & \msquark & \mchi   \\
             & $[\gevcc]$  &  $[\gevcc]$ &  [\gevcc]  &  &   &  [\gevcc]     &      [\gevcc]     \\

      \hline
      LV38  5/10/50ps    & 40    & 2000    & 2000 & 2.0   & 1200 &1300    & 38  \\
      LV98  5/10/50ps    & 100   & 2000    & 2000 & 2.0   & 1200 &1300    & 98  \\
      LV198 5/10/50ps    & 200   & 2000    & 2000 & 2.0   & 1200 &1300    & 198 \\ 
      \hline
     \end{tabular}
     }
    \label{tab:models-full}
  \end{center}
\end{table}

\clearpage


\addcontentsline{toc}{section}{References}
\setboolean{inbibliography}{true}
\bibliographystyle{LHCb}
\bibliography{main,LHCb-PAPER,LHCb-CONF,LHCb-DP,LHCb-TDR}

\newpage
\centerline{\large\bf LHCb collaboration}
\begin{flushleft}
\small
R.~Aaij$^{40}$,
B.~Adeva$^{39}$,
M.~Adinolfi$^{48}$,
Z.~Ajaltouni$^{5}$,
S.~Akar$^{6}$,
J.~Albrecht$^{10}$,
F.~Alessio$^{40}$,
M.~Alexander$^{53}$,
S.~Ali$^{43}$,
G.~Alkhazov$^{31}$
% ...

{\footnotesize \it
% ...
$ ^{5}$Clermont Universit{\'e}, Universit{\'e} Blaise Pascal, CNRS/IN2P3, LPC, Clermont-Ferrand, France\\
$ ^{6}$CPPM, Aix-Marseille Universit{\'e}, CNRS/IN2P3, Marseille, France\\
% ...
$ ^{10}$Fakult{\"a}t Physik, Technische Universit{\"a}t Dortmund, Dortmund, Germany\\
% ...
$ ^{31}$Petersburg Nuclear Physics Institute (PNPI), Gatchina, Russia\\
% ...
$ ^{39}$Universidad de Santiago de Compostela, Santiago de Compostela, Spain\\
$ ^{40}$European Organization for Nuclear Research (CERN), Geneva, Switzerland\\
% ...
$ ^{43}$Nikhef National Institute for Subatomic Physics, Amsterdam, The Netherlands\\
% ...
$ ^{48}$H.H. Wills Physics Laboratory, University of Bristol, Bristol, United Kingdom\\
% ...
$ ^{53}$School of Physics and Astronomy, University of Glasgow, Glasgow, United Kingdom\\
% ...
}
\end{flushleft}


\end{document}